# Spin-lattice instability to a fractional magnetization state in the spinel $HgCr_2O_4$

M. Matsuda[*], H. Ueda[†], A. Kikkawa[‡], Y. Tanaka[‡], K. Katsumata[‡], Y. Narumi[†], T. Inami[¶], Y. Ueda[†], S.-H. Lee[§]

[*] *Quantum Beam Science Directorate, Japan Atomic Energy Agency, Tokai, Ibaraki 319-1195, Japan*

[†] *The Institute for Solid State Physics, The University of Tokyo, Kashiwa, Chiba 277-8581, Japan*

[‡] *RIKEN SPring-8 Center, Harima Institute, Sayo, Hyogo 679-5148, Japan*

[¶] *Japan Atomic Energy Agency, Sayo, Hyogo 679-5148, Japan*

[§] *Department of Physics, University of Virginia, Charlottesville, Virginia 22904-4714, USA*

**Magnetic systems are fertile ground for the emergence of exotic states[1] when the magnetic interactions cannot be satisfied simultaneously due to the topology of the lattice – a situation known as geometrical frustration[2,3]. Spinels, $AB_2O_4$, can realize the most highly frustrated network of corner-sharing tetrahedra[1-3]. Several novel states have been discovered in spinels, such as composite spin clusters[1] and novel charge-ordered states[4-6]. Here we use neutron and synchrotron X-ray scattering to characterize the fractional magnetization state of $HgCr_2O_4$ under an external magnetic field, H. When the field is applied in its Néel ground state, a phase transition occurs at H ~ 10 Tesla at which each tetrahedron changes from a canted Néel state to a fractional spin state with the total spin, $S_{tet}$, of S/2 and the lattice undergoes orthorhombic to cubic symmetry change. Our results provide the**

**microscopic one-to-one correspondence between the spin state and the lattice distortion.**

Field-induced states in frustrated Heisenberg antiferromagnets are not easy to obtain experimentally. It is mainly because the fundamental spin degrees of freedom in this class of materials are nanoscale spin clusters with zero net moment[1,7,8] and therefore are highly insensitive to external magnetic field. The magnetic field has to be high enough so that the Zeeman energy can break the spin zero clusters. More often than not, the required field is very high for frustrated Heisenberg antiferromagnets and is not easily accessible for experimental investigations. For instance, $ZnCr_2O_4$, a highly studied and almost ideal model system for the pyrochlore lattice with isotropic S = 3/2 spins, has strong magnetic couplings between nearest neighboring $Cr^{3+}$ ions due to the direct overlap of their half-filled $t_{2g}$ orbitals[9,10] and thus requires a very high critical field. The strong coupling is evidenced by the large Curie-Weiss temperature, $\Theta_{CW}= JzS(S+1)/3k_B$ = -390 K [3,11], where J is the coupling constant, z the number of bondings, and $k_B$ the Boltzman constant. To break such strong magnetic couplings would require an external field of $H_c$ = gSJ (g ~ 2 is the gyromagnetic ratio) that is about 100 Telsa and is out of reach with the current technology.

Recently progress has been made in chemical synthesis to reduce J, which is a very strongly dependent function of $Cr^{3+}$-$Cr^{3+}$ distance, by replacing Zn with larger Cd and Hg, and thereby increasing the distance between the magnetic $Cr^{3+}$ ions [12,13]. Bulk magnetization measurements on $ACr_2O_4$ (A=Cd, Hg) under pulsed fields up to 47 T have revealed field-induced transitions to half-magnetization plateau phases in the low temperature Néel phase. For $CdCr_2O_4$ with $\Theta_{CW}$ = -88 K, the half-magnetization plateau phase occurs for H > 28 Tesla.[12] For $HgCr_2O_4$ with $\Theta_{CW}$ = -32 K, on the other hand, the half-magnetization plateau phase occurs for 10 Tesla < H < 28 Tesla and the full magnetization plateau phase for H > 40 Tesla.[13] The hysteretic nature of the field-induced transitions is observed with ramping down the field, indicating that the transitions are of first order.[13] The synthesis of $HgCr_2O_4$ is significant, even though it is in a polycrystalline form, because its critical field, ~10 T, can be achieved at neutron and X-ray scattering facilities and thereby it allows us to study in detail the nature of the half-magnetization state of the frustrated Heisenberg antiferromagnets. The key finding of our neutron and X-ray diffraction measurements on a polycrystalline sample of $HgCr_2O_4$ is that the stability of the half-magnetization state over 10.5 Tesla < H < 28 Tesla is due to a spin-lattice coupling that drives the spins and lattice to a field-induced state with cubic *P4₃32* symmetry. We explain the field-induced phase transition in terms of a simple modulation of the nearest neighbor spin interactions due to the lattice distortion.

Several different mechanisms to reduce frustration in the absence of an external magnetic field have been studied[11,13-22] -- for instance, the Jahn-Teller distortion in orbitally degenerate $AV_2O_4$ (A = $Zn^{18-20}$, $Cd^{21}$) and the spin-Peierls-like transition in $ACr_2O_4$ (A = $Zn^{11}$, $Cd^{22}$). These are evidenced by sudden drops in their bulk susceptibility measurements without field that involve cubic-to-tetragonal distortions. $HgCr_2O_4$ also shows similar behaviors at low temperatures to the other chromium spinels. Upon cooling without field, $HgCr_2O_4$ undergoes an orthorhombic distortion with *Fddd* symmetry at 6 K.[13] Elastic magnetic neutron scattering data obtained with zero field (blue circles in Fig. 1 (a)) shows that in the low temperature phase the spins order long range with two characteristic magnetic wave vectors $\mathbf{Q}_M = (1/2,0,1)$ and $(1,0,0)$. As listed in Table I, the relative intensities of different magnetic reflections tell us that the {1,0,0} peaks are due to a collinear spin structure where spins are parallel to the characteristic wave vector (the a-axis when we choose $\mathbf{Q}_M = (1,0,0)$) (Fig. 2 (b)). On the other hand, the {1/2,0,1} reflections are due to either a collinear or a non-collinear spin structure where spins are lying on the plane perpendicular to the doubling axis (the a-axis for $\mathbf{Q}_M = (1/2,0,1)$) (Fig. 2 (a)). Note that in both spin structures, each tetrahedron is made of two up-spins and two down-spins, and has total zero spin. Having the two different characteristic wave vectors suggests that the system may have two separate magnetic phases with $\mathbf{Q}_M = (1,0,0)$ and $(1/2,0,1)$, respectively. It is also possible that the Néel state of $HgCr_2O_4$ has a single magnetic phase where the *a*-component of the spins order with the $\mathbf{Q}_M = (1,0,0)$ whereas their *bc*-components with $(1/2,0,1)$. Frozen moment, <M>, was estimated from the comparison of the model and data to be 1.74(6) $\mu_B/Cr^{3+}$ which is less than the value of $gS\mu_B/Cr^{3+}$ with S=3/2. This indicates that strong frustrations exist even in the zero-field Néel phase.

Now let us turn to the half-magnetization plateau phase of $HgCr_2O_4$. With H = 13.5 Tesla that puts the system well in the half-magnetization state, we measured Q-dependence of elastic neutron scattering intensity at T = 3.2 and 10 K, below and above the transition. The 10.5 K data was subtracted from the 3.2 K data to obtain the magnetic contributions only, which is shown as red circles in Fig. 1 (a). Strikingly, the {1/2,0,1} magnetic peaks that were present at H = 0 have completely disappeared, while the {1,0,0} magnetic peaks became stronger. Furthermore, new magnetic peaks appeared at several nuclear Bragg reflection points such as (1,1,1), (1,3,1), and (2,2,2) but not at other nuclear Bragg reflection points such as (2,2,0). Note that the field-induced transition is rather sharp even though the sample is polycrystalline (Fig. 3 (a) and (b)). This suggests that the magnetic interactions are nearly isotropic even in the low temperature Néel phase, and regardless of the crystalline orientations, upon application of the external magnetic field, the spins reorient along the field. Such a weak anisotropy is due to the absence of orbital degeneracy in the system and the subsequent negligible spin-orbit couplings. The fact that the magnetization plateau phase over 10 Tesla < H < 28 Tesla has $<M> \sim \frac{gS}{2}\mu_B/Cr^{3+}$ suggests that each tetrahedron has three up spins and one down spin. With such tetrahedra, we can construct numerous spin structures for the lattice of corner-sharing tetrahedra with several different characteristic wave vectors. Among them, there are two nonequivalent spin configurations that would produce magnetic scattering at the {1,0,0} reflection points: one with rhombohedral $R\bar{3}m$ symmetry [Fig. 2 (c)] and the other with cubic $P4_332$ symmetry [Fig. 2 (d)]. In the magnetic structure with $R\bar{3}m$ symmetry, spins on the kagome planes (blue spheres in Fig. 2 (c)) along the <111> direction are up spins (parallel to H), whereas spins on the triangular planes (red spheres in Fig. 2 (c)) are down spins (antiparallel to H). On the other hand, in the structure with $P4_332$ symmetry, the nearest neighboring down-spins are separated by distances of $d_{\downarrow\downarrow}^{NN} = \{-1/4, 1/4, 1/2\}$, and the second nearest neighboring down-spins are separated by $d_{\downarrow\downarrow}^{2nd} = \{-1/2, 3/4, 1/4\}$. The $R\bar{3}m$ symmetry can be ruled out because it does not allow scattering at the observed (1,1,1) reflection, while it should produce magnetic scattering at the (220) reflection where no field induced magnetic signal has been observed (see Figs. 1 (a) and 3 (a)). On the other hand, all the observed field-induced magnetic reflections shown in Fig. 1 and 3 are allowed for *P4₃32* symmetry. A quantitative comparison between the experimental data and the model calculation is summarized in Table I, which shows the excellent agreement between the data and the *P4₃32* spin structure with the frozen moment of $<M> = 2.22(8) \mu_B/Cr^{3+}$.

Why is the half-magnetization phase stable over such a wide range of H? In order to address this issue, we performed synchrotron X-ray diffraction measurements with different fields and temperatures. Fig. 3 (c) shows that at 3.2 K (T < $T_N$), for H < 10 Tesla the (10,4,2) reflections splits into three peaks because as discussed earlier the lattice in the low field Néel phase is orthorhombic. For H > 10.5 Tesla, however, the peaks become one peak, which indicates that the external magnetic field changes the crystal structure from orthorhombic to cubic simultaneously as the system enters the half-magnetization plateau phase. Furthermore, as shown in Fig. 3 (d), the new nuclear peaks appeared at (4,4,1) and (5,2,2) reflections. This indicates that the crystal structure of the field-induced half-magnetization state has the same *P4₃32* symmetry as the magnetic structure. In the *Fddd* symmetry, $Cr^{3+}$ ions sit on crystallographically equivalent 16*d* symmetry sites that have inversion symmetry. On the other hand, in the *P4₃32* symmetry, Cr ions occupy two distinct symmetry sites, 4*b* and 12*d*. The 4*b* site is the common vertex of two neighboring tetrahedra each of which is composed of three *12d* and one *4b* sites. The tetrahedra are twisted and contracted along the local <111> direction that connects the *4b* site and the center of the *12d* triangles: the inversion symmetry is lost. Instead, the 4b site has a three-fold symmetry about a <111> axis and the 12b site has a two-fold symmetry along a <110> axis. Thus there are two different

Cr-Cr bond lengths in each Cr-tetrahedron: one between the $12d$ site ions and the other between the $4b$ and $12d$ site ions. By comparing an X-ray scattering peak corresponding to the {620} reflections with another peak corresponding to the {540}, {443}, and {621} reflections, we estimate the difference between the two bond lengths to be about 2%. Analysis of a series of chromium oxides indicates that the $J$ decreases when the distance between the Cr ions increases.[23] This implies that the distortion yields stronger AFM nearest neighbor (NN) couplings, $J_1$, between the $12d$ and $4b$ sites, and weaker AFM NN couplings, $J_1'$, between the $12d$ and $12d$ sites. This NN spatial exchange anisotropy will lift the degeneracy of the pyrochlore lattice composed of the excited tetrahedra and favor the spin structure that has the same cubic $P4_332$ symmetry as the lattice, i.e., the down-spin and the up-spin occupy the $4b$ and the $12d$ site, respectively (see Fig. 2 (d)). Therefore, we conclude that the half-magnetization state in $HgCr_2O_4$ is achieved and stabilized by the field-induced spin-lattice coupling that changes the system from the orthorhombic ($Fddd$) Néel state with total zero spin ($S_{tet} = 0$) tetrahedra to the cubic ($P4_332$) Néel state with $S_{tet} = S/2$ tetrahedra.

Previous studies have shown that spin-lattice coupling can lift the ground state degeneracy in frustrated magnets[11-17]. However, the detailed microscopic mechanism, i.e. the one-to-one correspondence between the selected Néel state and the local lattice distortion, has not yet been fully understood due to the complexity of the ensuing spin and lattice structures. Our neutron and synchrotron X-ray studies demonstrate for the first time how the spin-lattice instability leads to a particular lattice distortion and how the distortion favors the particular spin structure. These results allow us to test the validity of several theoretical models proposed towards understanding the spin-lattice coupling in frustrated magnets. For instance, two spin structures with $R\bar{3}m$ and $P4_332$ symmetries have been theoretically proposed for the field-induced half-magnetization state of $ACr_2O_4$[24-26]. The $R\bar{3}m$ spin structure proposed in Ref. [24] is preferred if the lattice undergoes a simple contraction along the <111> direction because such distortion will yield stronger AFM NN interactions between the kagomé (occupied by up spins) and triangular (down spins) planes, and weaker AFM NN interactions in the kagomé planes. On the other hand, the $P4_332$ spin structure proposed in Ref. [25] based on a purely magnetic mechanism is preferred if the local <111> lattice contraction is not uniform but propagates over the lattice with $P4_332$ symmetry, as have been observed in $HgCr_2O_4$. However, the fact that a lattice distortion occurs simultaneously with the magnetic symmetry breaking strongly indicates the necessity of a lattice distortion for the transition.[26]

## Methods

A polycrystalline sample of $HgCr_2O_4$ weighing about 2 g was pressed into a pellet and used for neutron and synchrotron X-ray scattering measurements. The neutron scattering measurements under magnetic field were carried out on the thermal neutron triple-axis spectrometer TAS2 at the JRR-3 reactor at Japan Atomic Energy Agency while measurements with no field were performed at the cold neutron triple-axis spectrometer SPINS at National Institute of Standards and Technology. The particular magnet used at TAS2 was a split-pair superconducting magnet cooled by cryocoolers and can reach fields up to 13.5 T. The fixed initial neutron energies were 14.7 meV and 5 meV on TAS2 and SPINS, respectively. Contamination from higher-order beams was effectively eliminated using the Polycrystalline Graphite and cooled Be filters at TAS2 and at SPINS, respectively. Synchrotron X-ray scattering was performed on BL19LXU at SPring-8 using an energy of 30 keV.

**Acknowledgements**

We thank S. Katano for use of the 13.5 T magnet, L. Balents, M. Gingras, K. Kakurai, D. I. Khomskii, K. Penc, H. Takagi for helpful discussions, and J.-H. Chung, Y. Shimojo for technical assistance. M.M. is supported by MEXT of Japan and JSPS. Activities at SPINS were partially supported by NSF. SHL is partially supported by NIST.

Correspondence and requests for materials should be addressed to S.H.L. (e-mail: shlee@virginia.edu).

**Figure captions**

**Fig. 1.** Q- and T-dependence of magnetic neutron scattering from a powder sample of $HgCr_2O_4$. (a) Magnetic neutron diffraction pattern obtained with H = 0 and 13.5 Tesla. The elastic intensity was measured below and above $T_N$ and the higher temperature data were subtracted from the lower temperature data to get rid of nonmagnetic contributions. (b) Temperature dependence of the peak intensities of the magnetic Bragg reflections at (0,1,1), and (3/2,0,1) measured with H = 0 and 13.5 Tesla. The data indicates that the Néel ordering temperature in the plateau phase increased to 7.5 K. Statistical errors were determined by the Poisson distribution, i.e., they are the square root of the intensities.

**Fig. 2.** Possible spin structures of the $Cr^{3+}$ ($3d^3$) (S=3/2) moments in $HgCr_2O_4$ (a),(b) at the ambient field and (c),(d) for the magnetization plateau state. Open and filled circles represent spin directions of being up and down, respectively. (a) For the $\mathbf{Q}_M$ = (1/2,0,1) and (b) for the $\mathbf{Q}_M$ = (1,0,0). (c) and (d) correspond to the structure for the $\mathbf{Q}_M$ = (1,0,0) with $R\bar{3}m$ and $P4_332$ symmetry, respectively. In (d), the distortion of each tetrahedron is exaggerated for display. The spins along the <110> chain directions in bc-planes in those models are $++--++--$, $+-+-+-+-$, and $+++-+++-$, for the models (a), (b) and (d), respectively.

**Fig. 3.** External magnetic field ($H$) effects on magnetic and nuclear Bragg scattering from a powder sample of $HgCr_2O_4$. (a),(b) $H$-dependence of the neutron scattering intensities at various reflections measured at 3.2 K, (a) at (2,2,0) and (1,1,1), and (B) at (3/2,0,1), (0,1,1) and (2,1,1). As $H$ increases up to 9 Tesla, the neutron diffraction pattern does not change (Fig. 1 (a)). However, the (111) Bragg intensity increases by $\sim 90 \pm 40$ counts/5 min. This increase is due to canting of spins along $H$, and we estimate the canting angle at 9 Tesla to be 17(4) degrees. The field-induced transition to the half-magnetization plateau phase occurs abruptly at H ~ 10 Tesla. (c),(d) Synchrotron X-ray diffraction data measured at several different temperature with different $H$s: (c) The nuclear (10,4,2) Bragg reflection. (d) The nuclear (4,4,1) and (522) reflections. Statistical errors were determined by the Poisson distribution.

**Table I.** The measured and calculated Q-integrated intensities of magnetic Bragg peaks for the HgCr$_2$O$_4$ pollycrystalline sample obtained with H = 0 and 13.5 Tesla. The intensities were normalized to the integrated intensity of the nuclear (2,2,2) Bragg reflection. The calculated intensities were obtained with the following frozen moments: For H = 0 data, <M$_{ab}$> = 1.02(2) μ$_B$/Cr$^{3+}$ lying on the bc-plane for the {1/2,0,1} reflections, and for the {1,0,0} reflections, <M$_c$> = 1.42(4) μ$_B$/Cr$^{3+}$ along the a-axis for the {1,0,0} reflections. Assuming that the zero-field Néel phase is a single phase, the magnitude of the frozen spin is <M> = 1.74(6) μ$_B$/Cr$^{3+}$. For H = 13.5 Tesla data, <M> = 2.22(8) μ$_B$/Cr$^{3+}$ along H.

| Q | H = 0 Tesla | | H = 13.5 Tesla | |
|---|---|---|---|---|
| | I$^{obs}$(Q) | I$^{cal}$(Q) | I$^{obs}$(Q) | I$^{cal}$(Q) |
| (1/2,1,0) | 0.028(3) | 0.029 | 0.0 | 0.0 |
| (0,1,1) | 0.049(3) | 0.050 | 0.16(2) | 0.15 |
| (1,1,1) | 0.0 | 0.0 | 0.07(1) | 0.06 |
| (3/2,0,1) | 0.090(1) | 0.086 | 0.0 | 0.0 |
| (1,0,2) | 0.058(6) | 0.057 | 0.17(2) | 0.19 |
| (-1/2,2,1) | 0.061(6) | 0.061 | 0.0 | 0.0 |
| (2,1,1) | 0.0 | 0.01 | 0.07(2) | 0.08 |
| (5/2,0,-1) | 0.035(6) | 0.038 | 0.0 | 0.0 |
| (0,3,1) | 0.0 | 0.016 | 0.03(2) | 0.04 |
| (2,2,2) | 0.0 | 0.0 | 0.04(2) | 0.04 |
| (3,0,2) | 0.0 | 0.008 | 0.04(1) | 0.05 |

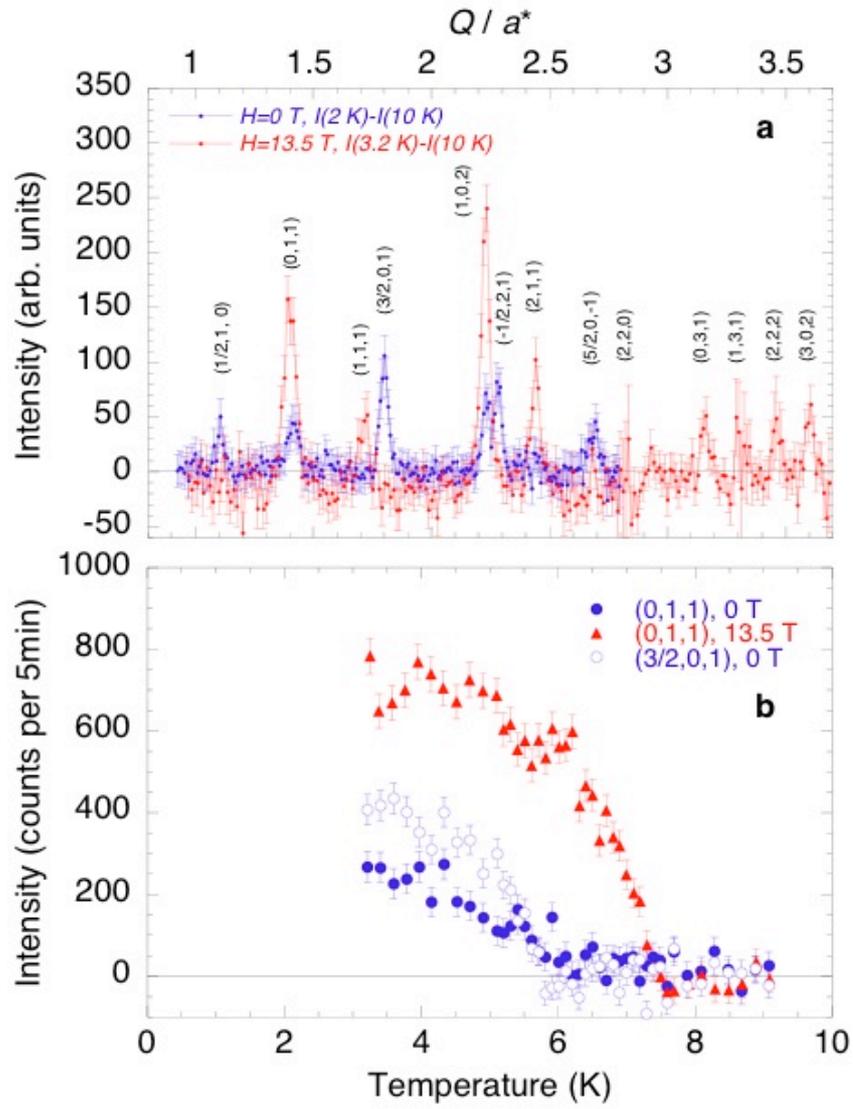

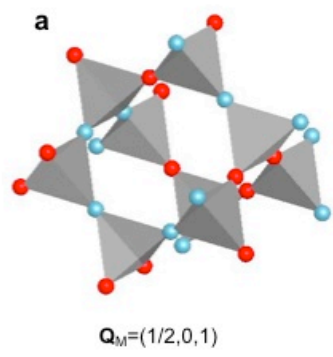

**a**

$\mathbf{Q}_M$=(1/2,0,1)

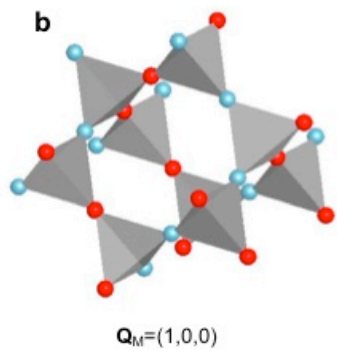

**b**

$\mathbf{Q}_M$=(1,0,0)

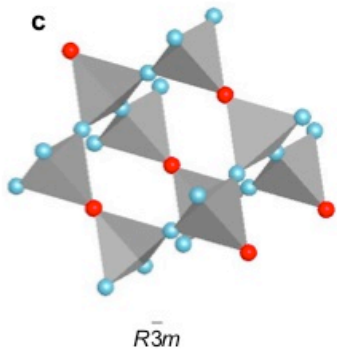

**c**

$R\bar{3}m$

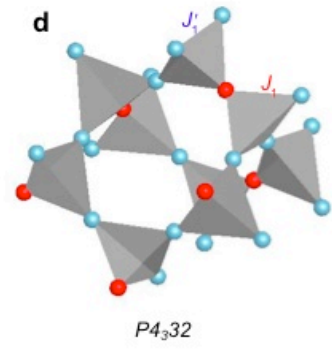

**d**

$P4_332$

$J_1'$  $J_1$

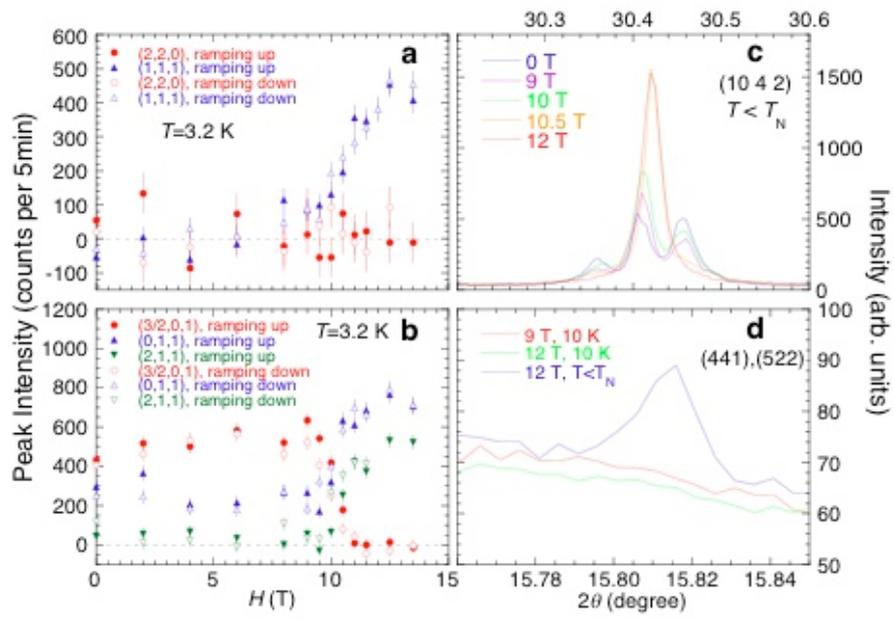